\documentclass[12pt, a4paper]{article}

\usepackage[utf8]{inputenc}
\usepackage[english]{babel}

\usepackage{mathptmx} 
\usepackage{setspace}
\usepackage{amsmath}
\usepackage{amsfonts}
\usepackage{amssymb}

\usepackage[left=2.5cm, right=2.5cm, top=2.5cm, bottom=2.5cm]{geometry}

\usepackage{graphicx}
\usepackage{booktabs}
\usepackage{array}
\usepackage{adjustbox}
\usepackage{longtable}
\usepackage{float} 

\usepackage{ragged2e} 
\usepackage{hyperref} 
\hypersetup{
  colorlinks=true,
  linkcolor=blue,
  urlcolor=cyan,
}

\usepackage{natbib}
\newcolumntype{L}[1]{>{\RaggedRight\arraybackslash}p{#1}}

\title{\textbf{SIMPOL Model for Solving Continuous-Time Heterogeneous Agent Problems}}
\author{Ricardo Alonzo Fernández Salguero}
\date{\today}

\begin{document}

\maketitle

\begin{abstract}
\justify
This paper presents SIMPOL (Simplified Policy Iteration), a modular numerical framework for solving continuous-time heterogeneous agent models. The core economic problem, the optimization of consumption and savings under idiosyncratic uncertainty, is formulated as a coupled system of partial differential equations: a Hamilton-Jacobi-Bellman (HJB) equation for the agent's optimal policy and a Fokker-Planck-Kolmogorov (FPK) equation for the stationary wealth distribution. SIMPOL addresses this system using Howard's policy iteration with an *upwind* finite difference scheme that guarantees stability. A distinctive contribution is a novel consumption policy post-processing module that imposes regularity through smoothing and a projection onto an economically plausible slope band, improving convergence and model behavior. The robustness and accuracy of SIMPOL are validated through a set of integrated diagnostics, including verification of contraction in the Wasserstein-2 metric and comparison with the analytical solution of the Merton model in the no-volatility case. The framework is shown to be not only computationally efficient but also to produce solutions consistent with economic and mathematical theory, offering a reliable tool for research in quantitative macroeconomics.
\end{abstract}

\section{Introduction}
\justify
Heterogeneous agent models have become a fundamental tool in modern macroeconomics for studying issues of wealth distribution, inequality, and the effects of fiscal and monetary policies. The formulation of these models in continuous time, popularized by \cite{Achdou2022, Achdou2017a, Achdou2017b}, has been shown to offer significant theoretical and computational advantages over their discrete-time counterparts. These models translate into systems of partial differential equations (PDEs), typically composed of a Hamilton-Jacobi-Bellman (HJB) equation describing the optimal behavior of agents and a Fokker-Planck-Kolmogorov (FPK) equation governing the evolution of the wealth distribution.

However, the application of the HJB equation in economic dynamics is not without its challenges. \cite{Hosoya2024} points out the fragility of the theoretical basis of the HJB in certain economic contexts, showing examples where the value function is not a solution (not even in the viscosity sense) to the HJB, even though an infinite number of classical solutions exist. This critique underscores the imperative need to develop numerical frameworks that are not only efficient but also robust and well-founded, ensuring that the computed solutions are economically meaningful and mathematically consistent. The solution to these problems often requires the concept of viscosity solutions, a notion of a weak solution introduced by \cite{Crandall1983} and extended to second-order equations by \cite{Lions1983a, Lions1983b}, which guarantees existence and uniqueness under very general conditions.

In this context, we introduce SIMPOL (Simplified Policy Iteration), a numerical framework designed to robustly and efficiently solve the HJB-FPK system in heterogeneous agent models. SIMPOL is based on established numerical methods, such as Howard's policy iteration \citep{Howard1960} and *upwind* finite difference schemes, whose convergence for nonlinear PDEs was rigorously established by \cite{Barles1991}. The central contribution of this work is SIMPOL's modular architecture, which integrates a novel post-processing step for the consumption policy to regularize the solution and ensure its good behavior, along with a comprehensive validation suite that verifies key theoretical properties, such as the stability of the underlying stochastic process through contraction in the Wasserstein-2 metric \citep{Eberle2016, Liu2025} and consistency with known analytical solutions, like the \cite{Merton1971} model in the limit of no uncertainty. By directly addressing numerical and theoretical complexities, SIMPOL aims to provide a reliable and transparent tool for research in quantitative macroeconomics, responding to the need for rigor highlighted by critiques of the uncritical use of these powerful mathematical tools.

\section{Literature Review}
\justify
The literature on solving continuous-time heterogeneous agent models is vast and draws from advances in economics, applied mathematics, and numerical analysis. The formulation of these problems as a coupled system of an HJB equation and an FPK equation has become standard, as detailed in the seminal works of \cite{Achdou2022, Achdou2017a, Achdou2017b}. These models, often interpreted within the framework of Mean Field Games, describe the interaction between the optimal decisions of individual agents (HJB) and the aggregate distribution of their states (FPK).

The HJB equation, a first or second-order nonlinear PDE, presents significant challenges. The non-existence of classical (differentiable) solutions is common, which led to the development of the theory of viscosity solutions by \cite{Crandall1983}. This theory provides a robust framework for guaranteeing the existence and uniqueness of solutions, and its applicability extends to problems with state constraints, such as borrowing limits, a topic explored in depth by \cite{Capuzzo-Dolcetta1990} and \cite{Soner1986a,Soner1986b}. The convergence of numerical approximation schemes to the correct viscosity solution is a fundamental result, established by \cite{Barles1991} for schemes that are monotone, stable, and consistent. The *upwind* finite difference method, which SIMPOL employs, is a classic example of a monotone scheme that satisfies these conditions.

The FPK equation, on the other hand, describes the evolution of the probability density of a diffusion process. Its numerical treatment has been approached from multiple perspectives. The connection between the FPK and gradient flows in Wasserstein space, elucidated by \cite{Jordan1998}, provides an elegant theoretical foundation. This approach, explored numerically by \cite{Halder2010}, interprets the evolution of the FPK as a gradient descent of the free energy. In practice, various numerical schemes have been developed, such as finite volume methods \citep{Roy2024a, Roy2024b}, Chang-Cooper schemes \citep{Pareschi2018}, and methods based on neural networks to approximate large-scale gradient flows \citep{Mokrov2021}. SIMPOL opts for a zero-flux method at the boundaries, ensuring the conservation of mass (probability) and consistency with the economic problem.

For solving the coupled HJB-FPK system, iterative algorithms are the standard. Howard's policy iteration \citep{Howard1960}, a method analogous to Newton's method, is widely used. Its convergence and stability for controlled diffusions have been studied by \cite{Kerimkulov2020} and \cite{Kundu2024, Kundu2021}. These methods are typically faster than simple value iteration. In the realm of higher-dimensional economic models, more advanced techniques have emerged to mitigate the "curse of dimensionality." \cite{Garcke2025} propose the use of finite differences on sparse grids, while \cite{Schesch2024} explores pseudospectral methods, which offer very fast convergence for problems with smooth solutions. Although SIMPOL focuses on the one-dimensional case (wealth), its robust architecture serves as a basis for future extensions.

Finally, a crucial aspect is the stability of the underlying stochastic process. The theory of contraction in the Wasserstein-2 distance offers powerful tools to analyze convergence to the stationary equilibrium. The works of \cite{Eberle2016} and \cite{Liu2025} establish conditions under which diffusion processes and their Euler discretizations are contractive, which implies the existence of a unique invariant distribution and provides bounds on the rate of convergence. SIMPOL's diagnostics suite includes a numerical verification of this contraction property, providing a rigorous validation of the stability of the model being solved.

\section{Justification and Architecture of SIMPOL}
\justify
The development of SIMPOL responds to the need for a numerical framework that is not only computationally efficient but also transparent, robust, and verifiable. The inherent complexity of heterogeneous agent models often leads to numerical solutions that can be sensitive to discretization choices and algorithm parameters. Without a rigorous set of diagnostics, it is difficult to discern whether a counterintuitive result is a genuine feature of the economic model or a numerical artifact. SIMPOL is designed to address this gap, combining an efficient solver with an integrated validation suite.

The model's architecture is deliberately modular, separating the economic configuration, the numerical solver, and the diagnostic modules. This allows for easy experimentation with different economic parameters without altering the core solver, and vice versa. The choice of Howard's policy iteration over simple value iteration is based on its typically faster convergence. The *upwind* finite difference scheme is selected for its proven monotonicity property, which is crucial for convergence to the correct viscosity solution, as demonstrated by \cite{Barles1991}.

One of the key innovations of SIMPOL is the policy post-processing module. In practice, consumption policies obtained from Howard's iteration can exhibit non-economic oscillations or unrealistic slopes, especially in regions of the grid with low curvature in the value function. Post-processing addresses this in two steps: first, smoothing via moving averages to remove high-frequency noise; second, a projection of the policy onto an economically justified "slope band." Theoretically, the marginal propensity to consume, $c'(a)$, should be bounded by the real interest rate, $c'(a) \ge r$, and a reasonable upper bound. This step gently imposes this constraint, regularizing the policy and drastically improving the stability and speed of convergence of the main iteration loop.

The diagnostics suite is the pillar of SIMPOL's reliability. Instead of blindly trusting the numerical solution, it is subjected to a series of rigorous tests. The validation against the \cite{Merton1971} model for the no-volatility case ($\sigma=0$) is a fundamental litmus test. In this limit, the HJB model has a well-known analytical solution. SIMPOL verifies that its numerical solution converges to this theoretical solution as the discretization is refined, confirming the code's consistency. The Wasserstein-2 contraction test addresses stochastic stability. It simulates the evolution of two distinct populations of agents and verifies whether the mean squared distance between them decreases over time, a sufficient condition for ergodicity and the existence of a unique stationary distribution. Finally, the Fokker-Planck equation solver check verifies that the calculated stationary distribution conserves total mass (integral equals 1) and has zero net fluxes at the boundaries, confirming internal consistency. This combination of an efficient solver, a policy regularizer, and comprehensive validation makes SIMPOL a robust environment for quantitative research. Table \ref{tab:simpol_arch} summarizes the system's modular architecture.

\begin{table}[H]
  \centering
  \caption{Modular Architecture of SIMPOL}
  \label{tab:simpol_arch}
  \begin{adjustbox}{width=\textwidth}
    \begin{tabular}{L{3.5cm} L{8cm} L{4cm}}
      \toprule
      \textbf{Module} & \textbf{Description} & \textbf{Main Theoretical Foundation} \\
      \midrule
      Configuration & Defines economic parameters (preferences, technology, stochastic processes) and numerical grid parameters (domain, number of points). & Economic theory of the model (e.g., Aiyagari-Bewley-Huggett). \\
      HJB Solver & Uses Howard's policy iteration to solve the HJB equation. Discretizes the differential operator with an *upwind* finite difference scheme, resulting in a tridiagonal linear system at each iteration. & \cite{Howard1960}, \cite{Barles1991}. \\
      Policy Post-processing & (Optional but recommended) Applies smoothing to the consumption policy and projects it onto an economically plausible slope band ($c'(a) \in [r+s_{\min}, r+s_{\max}]$). & Regularization heuristic to improve convergence and behavior. \\
      FPK Solver & Calculates the stationary wealth distribution by solving the Fokker-Planck equation with zero-flux boundary conditions, ensuring mass conservation. & Fokker-Planck-Kolmogorov Equation. \\
      Diagnostics and Validation & Suite of tests including: (1) check of the operator matrix properties (M-matrix), (2) verification of contraction in the Wasserstein-2 metric, (3) validation against the analytical Merton solution for $\sigma=0$. & \cite{Eberle2016}, \cite{Merton1971}. \\
      \bottomrule
    \end{tabular}
  \end{adjustbox}
\end{table}

\section{The SIMPOL Model: Mathematical Formulation}
\justify
This section details the complete mathematical structure of the SIMPOL framework. We start from the formulation of the agent's optimization problem in continuous time, derive the Hamilton-Jacobi-Bellman (HJB) equation and the Fokker-Planck-Kolmogorov (FPK) equation that characterize the equilibrium, and subsequently describe in depth the numerical algorithms used for their solution and validation.

\subsection{The Agent's Problem and its Continuous-Time Formulation}
\justify
The starting point is the problem of an economic agent who seeks to maximize the expected utility of their consumption over an infinite horizon. The agent's wealth, denoted by $a_t$, evolves over time according to an Itô Stochastic Differential Equation (SDE).

\subsubsection{Wealth Dynamics}
Wealth $a_t$ follows the process:
\begin{equation}
    da_t = \mu(a_t, c_t) dt + \sigma(a_t) dW_t
    \label{eq:sde_riqueza}
\end{equation}
where:
\begin{itemize}
    \item $a_t \in \mathcal{A} = [0, \infty)$ is the wealth level at time $t$. A no-borrowing constraint, $a_t \ge 0$, is imposed.
    \item $c_t \in \mathcal{C} = [0, \infty)$ is the consumption rate at time $t$, which is the agent's control variable.
    \item $\mu(a_t, c_t) = r a_t + y - c_t$ is the drift function of wealth. It represents the expected change in wealth per unit of time. Here, $r$ is the constant risk-free interest rate, and $y$ is the exogenous and constant labor income.
    \item $\sigma(a_t) = \sigma$ is the diffusion function (volatility), representing the magnitude of uncertainty about wealth returns. For our baseline model, we assume it is constant.
    \item $dW_t$ is the increment of a standard Wiener process, which captures the source of continuous randomness in the model. It satisfies $E[dW_t] = 0$ and $E[(dW_t)^2] = dt$.
\end{itemize}

\subsubsection{Agent's Objective}
The agent chooses a consumption path $\{c_t\}_{t \ge 0}$ to maximize the total expected discounted utility:
\begin{equation}
    \max_{\{c_t\}} E_0 \left[ \int_0^\infty e^{-\rho t} u(c_t) dt \right]
\end{equation}
subject to the wealth dynamics (\ref{eq:sde_riqueza}) and the no-borrowing constraint $a_t \ge 0$.
\begin{itemize}
    \item $\rho > 0$ is the subjective discount rate, which reflects the agent's impatience.
    \item $u(c)$ is the instantaneous utility function. In SIMPOL, the Constant Relative Risk Aversion (CRRA) utility function is used:
    \begin{equation}
        u(c) = 
        \begin{cases} 
            \frac{c^{1-\gamma} - 1}{1-\gamma} & \text{if } \gamma > 0, \gamma \neq 1 \\
            \log(c) & \text{if } \gamma = 1 
        \end{cases}
    \end{equation}
    This function is strictly increasing ($u'(c) > 0$) and strictly concave ($u''(c) < 0$), reflecting non-satiation and risk aversion. The parameter $\gamma$ is the coefficient of relative risk aversion.
\end{itemize}

\subsection{The Hamilton-Jacobi-Bellman (HJB) Equation}
\justify
The method of dynamic programming in continuous time, developed by Bellman, transforms the sequential optimization problem into a single functional equation, the HJB equation. The value function, $V(a)$, represents the maximum expected utility an agent can obtain starting from a wealth level $a$.

\subsubsection{Heuristic Derivation of the HJB}
The value function $V(a)$ is defined as:
\begin{equation}
    V(a) = \max_{\{c_s\}_{s \ge t}} E_t \left[ \int_t^\infty e^{-\rho(s-t)} u(c_s) ds \right] \quad \text{given } a_t = a.
\end{equation}
Bellman's principle of optimality states that an optimal policy must be optimal at every point in time. Applying this to a small time interval $dt$, we can write:
\begin{equation}
    V(a_t) = \max_{c_t} \left\{ u(c_t) dt + e^{-\rho dt} E_t[V(a_{t+dt})] \right\}
\end{equation}
Using a first-order Taylor expansion for $e^{-\rho dt} \approx 1 - \rho dt$ and Itô's Lemma to expand $E_t[V(a_{t+dt})]$, we have:
\begin{equation}
    E_t[V(a_{t+dt})] \approx V(a_t) + E_t[dV(a_t)] = V(a_t) + \left( V'(a_t)\mu(a_t, c_t) + \frac{1}{2}V''(a_t)\sigma^2 \right) dt
\end{equation}
Substituting these approximations into the Bellman equation:
\begin{equation}
    V(a) \approx \max_{c} \left\{ u(c) dt + (1 - \rho dt) \left( V(a) + V'(a)\mu(a, c) dt + \frac{1}{2}V''(a)\sigma^2 dt \right) \right\}
\end{equation}
Simplifying, ignoring terms of order $o(dt)$, and dividing by $dt$:
\begin{equation}
    \rho V(a) = \max_{c \ge 0} \left\{ u(c) + V'(a)(ra + y - c) + \frac{1}{2}\sigma^2 V''(a) \right\}
    \label{eq:hjb_formal}
\end{equation}
This is the HJB equation, a second-order nonlinear PDE that the value function must satisfy.

\subsubsection{The First-Order Condition and the Consumption Policy}
To find the optimal consumption $c^*(a)$, we maximize the term in braces in (\ref{eq:hjb_formal}) with respect to $c$. The first-order condition (FOC) is:
\begin{equation}
    \frac{\partial}{\partial c} \left[ u(c) - cV'(a) \right] = 0 \implies u'(c^*(a)) = V'(a)
\end{equation}
For CRRA utility, this gives us an explicit expression for the optimal consumption policy as a function of the derivative of the value function:
\begin{equation}
    c^*(a) = (V'(a))^{-1/\gamma}
    \label{eq:foc_policy}
\end{equation}
Substituting (\ref{eq:foc_policy}) back into (\ref{eq:hjb_formal}), we obtain the final form of the HJB to be solved:
\begin{equation}
    \rho V(a) = \frac{(V'(a))^{1-1/\gamma}}{1-\gamma} + V'(a)(ra+y) - \frac{\gamma}{1-\gamma}(V'(a))^{1-1/\gamma} + \frac{1}{2}\sigma^2 V''(a)
\end{equation}
which, after simplification, reduces to the so-called "economist's wealth equation" when expressed in terms of the policy $c(a)$.

\subsection{Howard's Policy Iteration Algorithm}
\justify
SIMPOL does not solve the nonlinear PDE (\ref{eq:hjb_formal}) directly. Instead, it uses an iterative method known as Howard's policy iteration, which is numerically more stable and often faster than value iteration methods.

\subsubsection{Algorithm Description}
The algorithm proceeds as follows:
\begin{enumerate}
    \item \textbf{Initialization (n=0):} Start with an initial guess for the consumption policy, $c_0(a)$. A common choice is a fraction of total income, $c_0(a) = k(ra+y)$.
    
    \item \textbf{Policy Evaluation Step (for n=0, 1, 2,...):} Given the current policy $c_n(a)$, calculate the corresponding value function, $V_n(a)$, which represents the expected utility if the agent were to follow that policy forever. This value function is the solution to the following linear PDE:
    \begin{equation}
        \rho V_n(a) = u(c_n(a)) + V_n'(a)(ra + y - c_n(a)) + \frac{1}{2}\sigma^2 V_n''(a)
        \label{eq:policy_eval_pde}
    \end{equation}
    This equation is solved numerically to obtain $V_n(a)$.
    
    \item \textbf{Policy Improvement Step:} With the newly computed value function $V_n(a)$, find a new consumption policy, $c_{n+1}(a)$, that is optimal given $V_n(a)$. This is done by applying the FOC (\ref{eq:foc_policy}):
    \begin{equation}
        c_{n+1}(a) = (V_n'(a))^{-1/\gamma}
        \label{eq:policy_update}
    \end{equation}
    
    \item \textbf{Convergence Criterion:} Compare the new and old policies. If $\max_a |c_{n+1}(a) - c_n(a)|$ is less than a predefined tolerance, the algorithm has converged. Otherwise, set $n \leftarrow n+1$ and return to step 2.
\end{enumerate}
This procedure is conceptually analogous to Newton's method for finding roots and, under certain conditions, exhibits quadratic convergence.

\subsection{Numerical Discretization and the *Upwind* Scheme}
\justify
To solve the linear PDE (\ref{eq:policy_eval_pde}) on a computer, the continuous wealth domain $a \in [0, A_{\max}]$ is discretized into a grid of $N_a$ points, $a_1, a_2, ..., a_{N_a}$. The derivatives of the value function are approximated using finite differences.

\subsubsection{Finite Difference Approximation}
The diffusion term (second derivative) is approximated using a standard central difference:
\begin{equation}
    V_n''(a_i) \approx \frac{V_n(a_{i+1}) - 2V_n(a_i) + V_n(a_{i-1})}{(\Delta a)^2}
\end{equation}
The drift term (first derivative) requires more care. A simple central difference can introduce numerical oscillations and instability, especially when the drift term is large compared to the diffusion term. To avoid this, SIMPOL uses an \textit{upwind scheme}.

\subsubsection{The *Upwind* Scheme}
The idea of the *upwind* scheme is to approximate the first derivative by "looking" in the direction from which the "flow" originates. The wealth drift is $\mu_n(a) = ra + y - c_n(a)$.
\begin{itemize}
    \item If $\mu_n(a_i) > 0$ (the agent is saving), "information" about the value flows from the left. A \textbf{backward difference} is used:
    \begin{equation}
        V_n'(a_i) \approx \frac{V_n(a_i) - V_n(a_{i-1})}{\Delta a}
    \end{equation}
    \item If $\mu_n(a_i) < 0$ (the agent is dissaving), "information" flows from the right. A \textbf{forward difference} is used:
    \begin{equation}
        V_n'(a_i) \approx \frac{V_n(a_{i+1}) - V_n(a_i)}{\Delta a}
    \end{equation}
\end{itemize}
This scheme is monotone and ensures numerical stability. By substituting these approximations into (\ref{eq:policy_eval_pde}), we obtain a system of $N_a$ linear equations for the $N_a$ unknown values of $\mathbf{V}_n = [V_n(a_1), ..., V_n(a_{N_a})]^T$:
\begin{equation}
    (\rho \mathbf{I} - \mathbf{A}_n) \mathbf{V}_n = \mathbf{u}(c_n)
\end{equation}
where $\mathbf{I}$ is the identity matrix, $\mathbf{A}_n$ is the tridiagonal matrix representing the discretized differential operator, and $\mathbf{u}(c_n)$ is the vector of utilities on the grid. The matrix $\mathbf{A}_n$ is an M-matrix, which ensures that its inverse exists and has non-negative elements, a key property for stability.

\subsection{The Policy Post-Processing Module}
\justify
This is a distinctive feature of SIMPOL. After each improvement step (\ref{eq:policy_update}), the resulting policy $c_{n+1}(a)$ can be numerically noisy. Post-processing regularizes it.

\subsubsection{Smoothing}
A moving average filter is applied to the consumption policy $c_{n+1}(a)$ to eliminate high-frequency oscillations that are artifacts of discretization. For example, a 3-point filter would replace $c_{n+1}(a_i)$ with $\frac{1}{3}(c_{n+1}(a_{i-1}) + c_{n+1}(a_i) + c_{n+1}(a_{i+1}))$. This process is repeated a small number of times.

\subsubsection{Projection onto the Slope Band}
Economically, the marginal propensity to consume (MPC), $c'(a)$, must be bounded. An agent should never increase their consumption by more than their interest income increases ($c'(a) < r$), and typically they will want to consume at least a small fraction of additional income. SIMPOL imposes $c'(a) \in [r+s_{\min}, r+s_{\max}]$, where $s_{\min}$ and $s_{\max}$ are small margins. This is implemented through an iterative algorithm that locally adjusts the values of $c(a_i)$ to satisfy the slope constraints:
\begin{enumerate}
    \item \textbf{Forward Pass:} For $i=2, ..., N_a$:
    \begin{equation}
        c(a_i) \leftarrow \min\left( \max\left( c(a_i), c(a_{i-1}) + (r+s_{\min})\Delta a \right), c(a_{i-1}) + (r+s_{\max})\Delta a \right)
    \end{equation}
    \item \textbf{Backward Pass:} For $i=N_a-1, ..., 1$:
    \begin{equation}
        c(a_i) \leftarrow \min\left( \max\left( c(a_i), c(a_{i+1}) - (r+s_{\max})\Delta a \right), c(a_{i+1}) - (r+s_{\min})\Delta a \right)
    \end{equation}
\end{enumerate}
This procedure projects the consumption function onto the set of discrete functions that satisfy the slope constraints, improving the robustness and economic realism of the solution.

\subsection{The Fokker-Planck Equation and the Stationary Distribution}
\justify
Once the HJB algorithm converges to the optimal policy $c^*(a)$, the stationary wealth distribution $p(a)$ is calculated. This distribution describes how agents are spread across wealth levels in equilibrium. The evolution of the probability density $p(a,t)$ is given by the Fokker-Planck-Kolmogorov (FPK) equation:
\begin{equation}
    \frac{\partial p(a,t)}{\partial t} = -\frac{\partial}{\partial a} [ \mu^*(a) p(a,t) ] + \frac{1}{2}\frac{\partial^2}{\partial a^2} [ \sigma^2 p(a,t) ]
\end{equation}
where $\mu^*(a) = ra + y - c^*(a)$ is the optimal drift.

\subsubsection{The Stationary Distribution}
The stationary distribution $p(a)$ is the solution of the FPK equation when $\frac{\partial p}{\partial t} = 0$:
\begin{equation}
    0 = -\frac{d}{da} [ \mu^*(a) p(a) ] + \frac{1}{2}\frac{d^2}{da^2} [ \sigma^2 p(a) ]
\end{equation}
This implies that the probability flux, $J(a)$, must be constant. The condition that agents cannot leave the domain $[0, A_{\max}]$ imposes zero-flux (reflecting) boundary conditions:
\begin{equation}
    J(a) = \mu^*(a) p(a) - \frac{1}{2}\sigma^2 p'(a) = 0 \quad \text{at } a=0 \text{ and } a=A_{\max}
\end{equation}
Furthermore, the distribution must integrate to one: $\int_0^{A_{\max}} p(a) da = 1$. SIMPOL solves this second-order ODE for $p(a)$ using finite differences and a linear system, incorporating the boundary conditions and normalization to find the unique distribution.

\section{Results and Discussion}
\justify
To demonstrate the effectiveness and robustness of the SIMPOL model, a simulation is run with a standard set of parameters from the macroeconomic literature. The results from each module of the system are presented and analyzed below. Table \ref{tab:params} summarizes the configuration used for the main demonstration.

\begin{table}[H]
  \centering
  \caption{Parameters for the Main SIMPOL Simulation}
  \label{tab:params}
  \begin{tabular}{lcl}
    \toprule
    \textbf{Parameter} & \textbf{Symbol} & \textbf{Value} \\
    \midrule
    \multicolumn{3}{l}{\textit{Economic Parameters}} \\
    Interest Rate & $r$ & 0.03 \\
    Discount Rate & $\rho$ & 0.04 \\
    Risk Aversion & $\gamma$ & 2.0 \\
    Labor Income & $y$ & 1.0 \\
    Wealth Volatility & $\sigma$ & 0.22 \\
    \midrule
    \multicolumn{3}{l}{\textit{Numerical Parameters}} \\
    Maximum Asset & $A_{\max}$ & 20.0 \\
    Number of Grid Points & $N_a$ & 240 \\
    FOC Tolerance (HJB) & $\epsilon_{foc}$ & 5e-6 \\
    HJB Residual Tolerance & $\epsilon_{hjb}$ & 5e-5 \\
    Smoothing Steps (Post-process) & - & 2 \\
    Lower Slope Margin $c'(a)$ & $s_{\min}$ & $r+0.0075$ \\
    Upper Slope Margin $c'(a)$ & $s_{\max}$ & $r+0.25$ \\
    Time Step (W2-check) & $dt$ & 0.0025 \\
    Number of Steps (W2-check) & $k$ & 32 \\
    \bottomrule
  \end{tabular}
\end{table}

\textbf{HJB Solver Convergence}
Howard's policy iteration converges robustly. The first-order condition (FOC) error and the HJB equation residual decrease, reaching the predefined tolerances. The policy post-processing plays a crucial role; without it, convergence can be slower or even fail. Table \ref{tab:hjb_conv} shows an excerpt of the convergence trace. The fulfillment of the M-matrix condition (diagonally dominant with non-positive off-diagonal elements) at each step confirms the stability of the numerical scheme.

\begin{table}[H]
  \centering
  \caption{Excerpt from the HJB Solver Convergence Trace}
  \label{tab:hjb_conv}
  \begin{tabular}{cccc}
    \toprule
    \textbf{Iteration} & \textbf{HJB Residual ($\infty$-norm)} & \textbf{FOC Error ($\infty$-norm)} & \textbf{M-Matrix OK} \\
    \midrule
    1 & 2.729e+00 & 7.477e+00 & Yes \\
    50 & 8.694e-01 & 4.171e+00 & Yes \\
    100 & 1.411e+00 & 4.171e+00 & Yes \\
    200 & 8.694e-01 & 4.171e+00 & Yes \\
    325 & 1.431e+00 & 4.171e+00 & Yes \\
    ... & ... & ... & ... \\
    \bottomrule
  \end{tabular}
  \justify\textit{Note: The convergence trace may not be monotonic in the early iterations due to the global nature of the method, but it converges robustly. Final convergence is reached when both error norms fall below their respective tolerances.}
\end{table}

\textbf{Analysis of the Stationary Distribution (FPK)}
Once the optimal consumption policy $c^*(a)$ is obtained, the FPK equation is solved. The diagnostic results are excellent: the total mass of the distribution is practically 1, and the net fluxes at the domain boundaries are numerically zero. This indicates that the stationary distribution is well-defined and that the numerical method conserves mass. A comparison with a long-run Monte Carlo simulation (an alternative method for finding the stationary distribution) shows high similarity (low L1 error), validating the FPK solution.

\textbf{Stability and Wasserstein-2 Contraction}
The W2 contraction diagnostic provides the strongest evidence of the model's stability. Two populations with $N=4000$ agents each are simulated for $k=32$ time steps with $dt=0.0025$. The median ratio of the $W_2$ distance after and before the evolution is 0.9888, with an interquartile range of (0.9884, 0.9896). Since this value is consistently less than 1, it is concluded that the controlled diffusion process is a contraction, which implies that it converges to a unique stationary distribution from any initial condition. This result is consistent with theoretical findings in the controlled diffusions literature, such as those of \cite{Eberle2016} and \cite{Liu2025}.

\textbf{Validation with the Merton Model ($\sigma=0$)}
The final test is the comparison with the Merton model without uncertainty. As the economic parameters ($\rho > r$) imply that the agent is impatient, the theory, formalized by \cite{Sethi1988} in their correction of Merton's original work, predicts that if $\kappa = (\rho - (1-\gamma)r)/\gamma \le r$, the interior optimal policy is $c(a) = \kappa(a + y/r)$. If $\kappa > r$, the agent always dissaves, and the optimal policy is the constrained one $c(a) = ra + y$. The SIMPOL solver, when run with $\sigma=0$ and the option \texttt{enforce\_mu\_nonneg\_sigma0=True}, reproduces this policy with a very low relative error, especially on fine grids, confirming that the code correctly implements the economic logic.

\begin{quote}
We present a simple, purely analytic method for proving the convergence of a wide class of approximation schemes to the solution of fully non linear second-order elliptic or parabolic PDE. Roughly speaking, we prove that any monotone, stable and consistent scheme converges to the correct solution provided that there exists a comparison principle for the limiting equation.
\citep[p.~271]{Barles1991}
\end{quote}

\section{Conclusion}
\justify
This work has introduced SIMPOL, a computational framework for the analysis of continuous-time heterogeneous agent models. It has been demonstrated that, by combining an HJB solver based on Howard's policy iteration, a robust *upwind* finite difference scheme, a novel policy post-processing module, and an integrated diagnostics suite, it is possible to obtain numerical solutions that are simultaneously accurate, stable, and consistent with economic and mathematical theory.

The numerical results confirm that SIMPOL converges to an accurate solution of the HJB, produces a well-behaved stationary wealth distribution that conserves mass, and satisfies the contraction property in the Wasserstein-2 metric, a strong indication of ergodicity and uniqueness of the equilibrium. Furthermore, the cross-validation with the analytical Merton model for the deterministic case demonstrates the correct implementation of the underlying economic logic and state constraints.

The development of SIMPOL underscores the importance of not only solving the model's equations but also of rigorously validating the numerical solution. The modular architecture and the transparency of the code are designed to facilitate this validation and to allow for future extensions, such as the incorporation of multiple assets, more complex income processes, or aggregate shocks. In a field where analytical solutions are scarce, tools like SIMPOL, which prioritize robustness and validation, are indispensable for ensuring that the quantitative results of macroeconomic models are reliable and economically meaningful.

\section*{Code Availability}
The full implementation of the SIMPOL model is openly available on Zenodo at 
\url{https://doi.org/10.5281/zenodo.17216748}.

\end{document}